\documentclass[12pt]{article}
\usepackage{wallpaper}
\usepackage[margin=1in, paperwidth=8.5in, paperheight=11in]{geometry}
\usepackage{amsmath}
\usepackage{graphicx}%
\usepackage{amsfonts}%
\usepackage{amssymb}
\usepackage{color}
\usepackage{float,subfigure}
\usepackage[numbers,sort&compress]{natbib}
\usepackage{cite}
\usepackage{setspace}
\usepackage{comment}
\usepackage{enumerate}
\usepackage{verbatim}
\usepackage{rotating}
\usepackage{microtype}
\usepackage[normalem]{ulem}
\usepackage{url}
\usepackage[hidelinks]{hyperref} 
\usepackage{cancel}
\usepackage{animate}
\usepackage{mathtools}
\usepackage{epsfig}
\usepackage{epstopdf}
\usepackage{multirow}
\usepackage{amsthm}
\usepackage{caption}
\subcapraggedrighttrue
\usepackage{cleveref}
\hypersetup{pdfpagemode=UseNone}
\theoremstyle{plain}
\newtheorem*{thm*}{Theorem} 

\theoremstyle{definition}

\newcommand{\appropto}{\mathrel{\vcenter{
  \offinterlineskip\halign{\hfil$##$\cr
    \propto\cr\noalign{\kern2pt}\sim\cr\noalign{\kern-2pt}}}}}

\newcommand{\bbeta}{ \ensuremath{\boldsymbol{\beta}}}

\newcommand{\btheta}{ \mbox{\boldmath $ \theta $} }
\newcommand{\sig}{ \ensuremath{\sigma}}

\newcommand{\bx}{ {\bf x} }

\newcommand{\by}{ {\bf y} }

\newcommand{\bz}{ {\bf z} }

\newcommand{\texp}  { \mbox{\text exp}}
\newcommand{\tlog}  { \mbox{\text log}}
\newcommand{\tlogit} { \mbox{\text logit}}

\newcommand{\Var}{\mbox{Var}}

\newcommand{\given}{\,\vert\,}

\newcommand{\CAR}{\mbox{$\text{CAR}$}}

\newcommand{\tbeta}{\mbox{$\text{Beta}$}}

\newcommand{\IG}{\mbox{$\text{IG}$}}

\newcommand{\Pois}{\mbox{$\text{Pois}$}}
\newcommand{\Bin}{\mbox{$\text{Bin}$}}

\newcommand{\N}{\mbox{Norm}}

\newcommand{\LogitN}{\mbox{LogitNorm}}
\newcommand{\Unif}{\mbox{Unif}}

\begin{document}


\thispagestyle{empty}
\setcounter{page}{0}
\singlespacing
\begin{center}
{\Large \textbf{Geographic and Racial Disparities in the Incidence of Low Birthweight in Pennsylvania}} %

\bigskip

\textbf{Guangzi Song, Loni Philip Tabb, and Harrison Quick$^{*}$}\\ 
 Department of Epidemiology and Biostatistics, Drexel University, Philadelphia, PA 19104\\
$^{*}$ \emph{email:} hsq23@drexel.edu

\end{center}

\textsc{Summary.}
Babies born with low and very low birthweights --- i.e., birthweights below 2,500 and 1,500 grams, respectively --- have an increased risk of complications compared to other babies, and the proportion of babies with a low birthweight is a common metric used when evaluating public health in a population. While many factors increase the risk of a baby having a low birthweight, many can be linked to the mother’s socioeconomic status, which in turn contributes to large racial disparities in the incidence of low weight births. Here, we employ Bayesian statistical models to analyze the proportion of babies with low birthweight in Pennsylvania counties by race/ethnicity. Due to the small number of births --- and low weight births --- in many Pennsylvania counties when stratified by race/ethnicity, our methods must walk a fine line. On one hand, leveraging spatial structure can help improve the precision of our estimates. On the other hand, we must be cautious to avoid letting the model overwhelm the information in the data and produce spurious conclusions. As such, we first develop a framework by which we can measure (and control) the informativeness of our spatial model. After demonstrating the properties of our framework via simulation, we analyze the low birthweight data from Pennsylvania and examine the extent to which the commonly used conditional autoregressive model can lead to oversmoothing.  We then reanalyze the data using our proposed framework and highlight its ability to detect (or not detect) evidence of racial disparities in the incidence of low birthweight.


\newpage

\doublespacing
\section{Introduction}
The incidence of low and very low birthweight --- i.e., birthweights below 2,500 and 1,500 grams, respectively --- are important indicators of public health. Not only is low birthweight a leading risk factor for infant mortality, it has also been linked to an increased risk of serious long-term disabilities \citep{Hack} and onset of chronic diseases in adulthood \citep{collins,Barker,Rich}. Worldwide, it is estimated that 15--20\% of all births --- and more than 20 million births per year --- have low birthweight \citep{who}. While the incidence of low birthweight in the United States (U.S.) is much lower, 2017 marked the highest level in over 20 years (8.27\%) --- a value that was subsequently exceeded in both 2018 and 2019 \citep{births:2019}.  Of course, it is also true that the incidence of low birthweight is not equal for all populations.  
Reports from the Centers for Disease Control and Prevention (CDC) indicate increasing racial disparities in low birthweight from 2015 to 2019, with 
the incidence of low birthweight for black mothers more than twice 
that of white mothers \citep{births:2019}.  Furthermore, geographic disparities in the incidence of low birthweight are also observed, with rates in Mississippi (12.3\%) being nearly double those in Alaska (6.3\%) \citep{births:2019}.

While the topics of racial and geographic disparities are important in their own right, it is also important to study geographic trends in racial disparities, as this can shed light on regions in which all race/ethnicities have equal health outcomes and regions where some racial/ethnic populations are being left behind.  For instance, \citet{goldfarb} assessed the incidence of low birthweight in 400 of the largest U.S.\ counties from 2003--2013 and found that only four counties demonstrated improvement toward racial equality via improvement in the incidence of low birthweight for black mothers (rather than \emph{worsening} trends for white mothers). Unfortunately, such studies of low birthweight are plagued by two key issues.  First and foremost, while the CDC's Wide-ranging Online Data for Epidemiologic Research (WONDER) is a great resource for accessing \emph{mortality} data, data available from CDC WONDER pertaining to births is limited to counties with population sizes larger than 100,000.  Not only does this represent fewer than 20\% of U.S.\ counties, but it also precludes inference on urban/rural disparities in the incidence of low birthweight. The second key challenge encountered when investigating geographic trends in racial disparities in low birthweight is that even when access to detailed county-level data is available via state-specific web portals --- e.g., the Commonwealth of Pennsylvania's Enterprise Data Dissemination Informatics Exchange (EDDIE) system --- many parts of the country have small racial/ethnic minority populations. Because counties in these areas are more likely to have few births to non-white mothers, estimates of the incidence rates of low birthweight births for minority populations may be unreliable or otherwise lack the level of precision required for inference.

The objective of this paper is to develop a modeling framework that offers a compromise between increasing the precision of estimates of rates such as the incidence of low birthweight while protecting against oversmoothing --- i.e., when the contribution of the model overwhelms that of the data and distorts the underlying trends.  As a starting point, we consider the use of Bayesian methods that incorporate spatial structure into the model, such as the conditional autoregressive (CAR) model framework of \citet{bym}.  Aside from its use in a variety of applications \citep{waller:carlin,gelfand:mcar,hcar,b-r:2015}, recent work by \citet{bym:info} has illustrated the extent to which the CAR model framework can oversmooth estimates by virtue of putting excessive weight on the model's spatial structure, thereby overwhelming the contribution of the data from individual regions.  For instance,
\citet{bym:info} analyzed 
county-level, heart disease-related death data and estimated the contribution of the model as being equivalent to 36 additional deaths occurring in each county, despite many counties having fewer than 10 observed events.

In this paper, we extend the work of \citet{bym:info} from the Poisson-distributed count setting to the case of binomially distributed data for the purpose of investigating racial disparities in the incidence of low birthweight in Pennsylvania counties.  In Section~\ref{sec:data}, we describe the data used in this analysis.  Section~\ref{sec:methods} provides a brief overview of the approach of \citet{bym:info} for quantifying (and restricting) the informativeness of the \citeauthor{bym} CAR framework and describes its extension to binomial data.  We then evaluate the properties of our approach via simulation in Section~\ref{sec:sim} prior to implementing our approach to investigate trends in low birthweight in Section~\ref{sec:analysis}.  We then conclude with a discussion in Section~\ref{sec:disc}.  

\section{Pennsylvania Birth Data}\label{sec:data}
We use county-level birth data in Pennsylvania for 2019 obtained from the Pennsylvania Department of Health's web-based EDDIE system \citep{eddie}. 
Via the EDDIE system, users can access data such as death and birth statistics from the 67 counties of Pennsylvania stratified by year and demographics such as age, race/ethnicity, and sex.  Unlike CDC WONDER, data obtained from the EDDIE system are free of suppression regardless of the number of events or the underlying population size.  For this analysis, we consider all live births in which the mother's race/ethnicity was categorized as white, black, Hispanic, or Asian/Pacific Islander (Asian),
with the caveat that these groups are \emph{not} mutually exclusive in the EDDIE system (e.g., mothers categorized as ``white'' include Hispanic and non-Hispanic white mothers).

Table~\ref{tab:descriptive} provides an overview of the Pennsylvania birth data used in this analysis, along with data from mothers of all races (including those not analyzed here).  First, we note that the overall and race-specific rates of low and very low birthweight are largely consistent with those observed at the national level \citep{births:2019}.  Nevertheless, Asian and Hispanic mothers have rates that are 25\% greater than for white mothers and black mothers have rates that are more than double than their white counterparts, with more extreme disparities observed in the incidence of very low birthweight.  In addition to racial disparities, this table also highlights the challenge of estimating rates at the county-level from these data, as each of the non-white populations have fewer than ten \emph{total} births in over one-third of counties and fewer than ten \emph{low weight} births in over 60\% of counties.  These challenges are only exacerbated when investigating rates of very low birthweight, with over 47\% of counties having fewer than ten very low weight births across all race/ethnicities.

\begin{table}[t]
\begin{center}
\resizebox{\textwidth}{!}{%
\begin{tabular}{ |c |c |c |c |c |c|}
\hline
{} & White & Black & Asian & Hispanic & All Races \\
\hline
Births & 92,702 & 18,798 & 6,237 & 16,657 & 133,628 \\  
Low weight births & 6,489 (7.0\%)& 2,707 (14.4\%) & 548 (8.8\%) & 1,508 (9.1\%)& 11,311 (8.5\%)\\ 
Very low weight births & 964 (1.0\%) & 573 (3.0\%) & 80 (1.3\%) & 275 (1.7\%) & 1931 (1.4\%)\\
Counties $<$ 10 births & 0 (0\%) & 31 (46.3\%) & 38 (56.7\%) & 24 (35.8\%) & 0 (0\%)\\ 
Counties $<$ 10 l.w.\ births & 4 (6.0\%) & 44 (65.7\%) & 55 (82.1\%) & 47 (70.1\%) & 4 (6.0\%)\\
Counties $<$ 10 v.l.w.\ births & 35 (52.2\%) & 58 (86.6\%) & 64 (95.5\%) & 57 (85.1\%) & 32 (47.8\%)\\
\hline
\end{tabular} }
\end{center}
\caption{Descriptive statistics of the Pennsylvania birth data described in Section~\ref{sec:data}.}
\label{tab:descriptive}
\end{table}

\section{Methods}\label{sec:methods}
Before discussing the specifics of our framework, we will first motivate the assumption that the number of low weight births should be treated as being binomially distributed.  To do so, we consider the guidance of \citet{brillinger}, which recommends modeling vital statistics such as the number of births and deaths as Poisson random variables.  Because our interest is to make inference on the proportion of births with low birthweight, we could first let $y_i$ and $y_i^*$ denote the number of low and non-low weight births in region $i$ from a total population of size $N_i$ and assume that $y_i\given \lambda_i \sim \Pois\left(N_i\lambda_i\right)$ and $y_i^*\given \lambda_i^* \sim \Pois\left(N_i\lambda_i^*\right)$, where $\lambda_i$ and $\lambda_i^*$ represent the \emph{per capita} rate of low and non-low weight births in each region.  Then, if we let $n_i = y_i+y_i^*$ denote the total number of births in region $i$, it would follow that $y_i\given n_i,\pi_i \sim \Bin\left(n_i,\pi_i\right)$, where $\pi_i=\lambda_i\slash \left(\lambda_i+\lambda_i^*\right)$; a similar logic will be used with respect to our analysis of the incidence of \emph{very} low weight births.  While the data described in Section~\ref{sec:data} and analyzed in Section~\ref{sec:analysis} are indexed by both county and race/ethnicity, in this section we simply index our variables by geographic region for the sake of simplicity.  

Because we aim to use a Bayesian framework to study the
county-level incidence of low weight births, we must select a prior distribution for each of our $\pi_i$ parameters.  A natural --- albeit simplistic --- starting point would be to consider the conjugate prior setting in which $\pi_i \sim \tbeta\left(a_i,b_i\right)$, where $a_i>0$ and $b_i>0$ denote the model's hyperparameters.  Not only does this choice of prior lead to a closed form expression for the posterior distribution,
\begin{align}
\pi_i\given y_i,n_i \sim \tbeta\left(y_i+a_i, n_i-y_i+b_i\right), \label{eq:betabin}
\end{align}
but it also produces convenient interpretations of our hyperparameters: i.e., that $a_i$ and $b_i$ correspond to the prior number of low and non-low weight births, respectively, and thus that $a_i+b_i$ represents the prior number of births.  This is beneficial for two key reasons.  First and foremost, it allows us to quickly identify the value that our prior is smoothing rates toward --- i.e., $E\left[\pi_i\given a_i,b_i\right] = a_i\slash \left(a_i+b_i\right)=\pi_{i0}$.  In addition, it allows it quantify the amount of information contributed by the prior --- i.e., we can compare the value of $y_i$ to $a_i$ and the value of $n_i$ to $a_i+b_i$ to ensure that our prior distribution is not unduly (or \emph{unexpectedly}) overwhelming the contribution of the data in~\eqref{eq:betabin}.

While the conjugate prior specification is well suited for explaining the informativeness of the model, the use of such models is not common in disease mapping settings.  Instead, it's more common to use a generalized linear mixed model framework --- i.e., $g\left(\pi_i\right) \sim \N\left(\bx_i^T\bbeta,\sig_i^2\right)$, where $g(\cdot)$ denotes an appropriate link function (e.g., $g\left(\pi_i\right) = \log \pi_i\slash\left(1-\pi_i\right)$). The trade-off of using such models, however, is that measuring their informativeness is not straightforward.  As such, our plan for the remainder of this section is as follows.  First, we follow the approach of \citet{bym:info} to approximate the informativeness of a $\pi_i\sim \LogitN\left(\mu_i,\sig_i^2\right)$ prior specification in Section~\ref{sec:logit}.  We then briefly describe how this approximation can be extended to the CAR model framework of \citet{bym} in Section~\ref{sec:bym}.

\subsection{Extension to logitnormal prior}\label{sec:logit}

While the model in~\eqref{eq:betabin} is transparent in what it smooths values toward and the strength of this smoothing, 
specifying the $a_i$ and $b_i$ parameters to accommodate features such as covariate information and/or complex dependence structures is not straightforward.  In contrast, the drawback of the logitnormal specification mentioned above is that its posterior distribution,
\begin{align}
\pi_i \given y_i, \mu_i, \sig_i^2 &\propto \Bin(y_i\given n_i, \pi_i) \times \LogitN(\pi_i\given \mu_i, \sig_i^2)\notag\\
&\propto \pi_i^{y_i} (1-\pi_i)^{n_i-y_i}\times \texp\left[-\frac{\left(\tlog\left(\pi_i/\left(1-\pi_i\right)\right)-\mu_i\right)^2}{2\sig_i^2}\right]
\label{eq:binlogit}
\end{align}
is not of a known form and thus lacks convenient expressions for quantities like its mean and variance.  As such, our strategy here is to follow the approach of \citet{bym:info} and construct priors of the form $\pi_i \sim \LogitN\left(\mu_i,\sig_i^2\right)$ that yield approximately the same inference as a hypothetical $\pi_i \sim \tbeta\left(a_i,b_i\right)$ prior.  More specifically, if we can write $\mu_i$ and $\sig_i^2$ as functions of $a_i$ and $b_i$ and vice versa, then we can \emph{quantify} the informativeness of the logitnormal prior or impose \emph{restrictions} on the bivariate distribution for $\left(\mu_i,\sig_i^2\right)$ such that the logitnormal prior can be no more (or \emph{no less}) informative than its analogous beta prior.  Unlike the lognormal distribution discussed in \citet{bym:info}, a challenge we face here is that the logitnormal distribution itself does not have analytical solutions for its mean and variance.  Thus rather than \emph{directly} equating the mean and variance of the beta and logitnormal distributions, we apply the delta method to approximate the mean and variance of a logitnormally distributed random variable by first supposing that $\theta_i\sim\N\left(\mu_i,\sig_i^2\right)$ and
$\pi_i = h(\theta_i)=\tlogit^{-1}\theta_i = \frac{\exp\theta_i}{1+\exp\theta_i}.$  Since $h(\theta)$ is a differentiable function about $\theta$, 
the delta method can be used to produce the approximation that $\pi_i \given \mu_i,\sig_i^2 \sim \N\left(E\left[\pi_i\given \mu_i,\sigma_i^2\right], \Var\left[\pi_i\given \mu_i,\sigma_i^2\right]\right)$, where
\begin{align}
E\left[\pi_i\given \mu_i,\sigma_i^2\right] &\approx h(\mu_i) =\frac{\exp\mu_i}{1+\exp\mu_i}\notag\\
\Var\left[\pi_i\given \mu_i,\sigma_i^2\right] &\approx \left[\frac{\partial^2}{\partial\mu_i^2}h(\mu_i)\right]^2\times\Var\left[\theta_i\given \mu_i,\sigma_i^2\right]=\frac{\sigma_i^2\left(\exp\mu_i\right)^2}{\left(1+\exp\mu_i\right)^4},\label{eq:delta}
\end{align}
as shown in Appendix~A.  Finally, we follow the approach of
\citet{bym:info} and equate the mean and variance of beta distribution with their respective expressions in \eqref{eq:delta}, which leads to
\begin{align}
E\left[\pi_i\given a_i,b_i\right] &\approx E\left[\pi_i\given \mu_i,\sig_i^2\right] &\implies&&  \frac{a_i}{a_i+b_i} &\approx \frac{\exp\mu_i}{1+\exp\mu_i};\notag \\
\Var\left[\pi_i\given a_i,b_i\right] &\approx \Var\left[\pi_i\given \mu_i,\sig_i^2\right] &\implies&&  \frac{a_ib_i}{(a_i+b_i)^2(a_i+b_i+1)} &\approx \frac{\sig_i^2\left(\exp\mu_i\right)^2}{\left(1+\exp\mu_i\right)^4}.
\label{eq:dev}
\end{align}
This allows us to express the logitnormal distribution parameters, $\mu_i$ and $\sigma^2_i$, in the form of the parameters $a_i$ and $b_i$ from the beta prior distribution,
\begin{align}
\mu_i&=\log \frac{a_i}{b_i} \hspace{0.1in}\mbox{and}\hspace{0.1in} \sigma_i^2 = \frac{\left(a_i+b_i\right)^2}{a_i b_i\left(a_i+b_i+1\right)},\label{eq:res}
\end{align}
and quantify the informativeness of the logitnormal prior as
\begin{align}
\widehat{a}_i&=\frac{1+\exp\mu_i}{\sigma_i^2}-\frac{\exp\mu_i}{1+\exp\mu_i}.\label{eq:a}
\end{align}

\begin{figure}[t]
    \begin{center}
        \subfigure[$E\left(\pi_i\given a_i,b_i\right)=1\slash 100$]{\includegraphics[width=.45\textwidth]{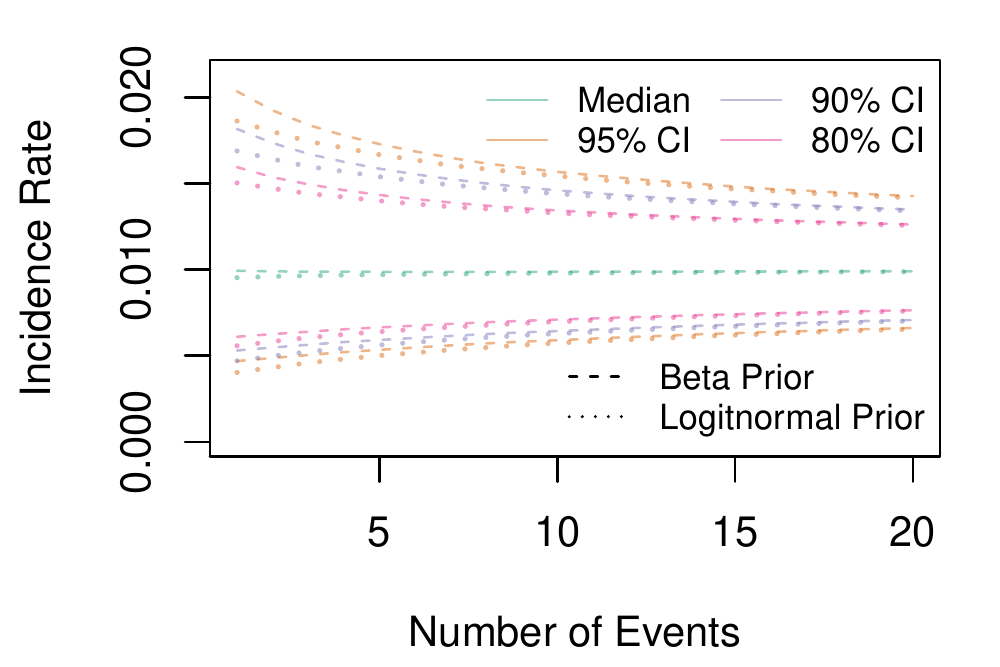}\label{fig:fig11}}
        \subfigure[$E\left(\pi_i\given a_i,b_i\right)=1\slash 3$]{\includegraphics[width=.45\textwidth]{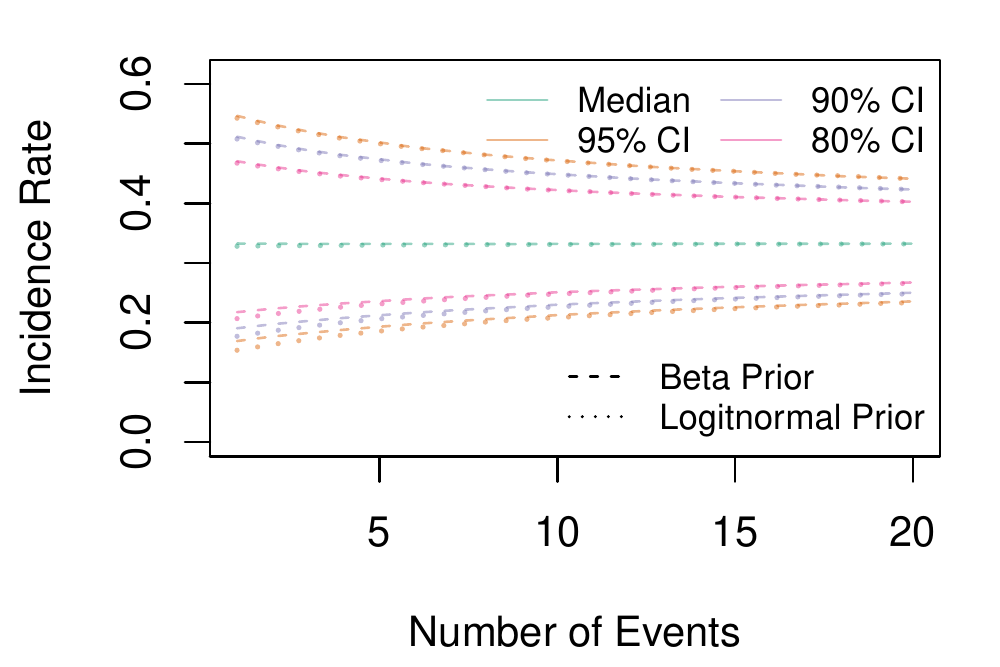}\label{fig:fig12}}
    \end{center}
    \caption{Comparison of quantiles of the posterior distribution for $\pi_i$ for beta and logitnormal prior distributions for various levels of events for the approximation in~\eqref{eq:res}. Panel~(a) displays results for $E\left[\pi_i\given a_i,b_i\right]=1\slash 100$, while Panel~(b) displays results for $E\left[\pi_i\given a_i,b_i\right]=1\slash 3$.}
	\label{fig:fig1}
\end{figure}

Figure~\ref{fig:fig1} illustrates the accuracy of the logitnormal approximation in~\eqref{eq:res} when $a_i=6$ under two scenarios: when an event is rare ($E\left[\pi_i\given a_i,b_i\right]=1\slash 100$; Figure~\ref{fig:fig11}) and when an event is quite common ($E\left[\pi_i\given a_i,b_i\right]=1\slash 3$; Figure~\ref{fig:fig12}).
Specifically, we compare quantiles of the posterior distribution for $\pi_i$ from a beta prior specification evaluated at various levels of event counts, $y_i$, such that $\pi_0=a_i/(a_i+b_i)=y_i/n_i$ to a logitnormal prior specification based on the approximation in~\eqref{eq:res}.
Here, we see that while the two distributions are not \emph{identical} --- with quantiles from the logitnormal prior often less than those from the beta prior and with larger deviations present when the number of events is low --- the two posterior distributions are quite similar. Based on these results, we claim that the prior $\pi_i\sim \LogitN\left(\mu_i,\sigma_i^2\right)$ is approximately as informative as $\pi_i \sim \tbeta\left(a_i,b_i\right)$ when $\mu_i$ and $\sigma_i^2$ are defined as shown in~\eqref{eq:res}. Evaluating this claim will be the focus of the simulation study in Section~\ref{sec:sim}.

\subsection{Extension to CAR model}\label{sec:bym}
While the approximation described in Section~\ref{sec:logit} extended the concept of prior informativeness from the beta distribution to the logitnormal, quantifying the informativeness of \emph{correlated} prior specifications requires knowledge of the \emph{conditional} distribution of $\pi_i$, as demonstrated by \citet{bym:info}. In the context of the CAR model of \citet{bym}, we assume logit$(\pi_i)=\theta_i$ for $i=1,\dots, I$ has a prior specification of the form:
\begin{align}
\theta_i \given \bbeta,\bz,\sig^2 &\sim \N\left(\bx_i^T\bbeta + z_i,\sig^2\right), \label{eq:bym}
\end{align}
where $\bx_i$ is a $p\times 1$ vector of covariates for region $i$ with corresponding regression coefficients, $\bbeta$, and $z_i$ is a spatial random effect for region $i$ such that $\bz \sim \CAR(\tau^2)$, which implies
\begin{align}
z_i \given \bz_{(i)},\tau^2 &\sim \N\left(\sum_{j\sim i} z_j \slash m_i, \tau^2\slash m_i\right), \label{eq:car}
\end{align}
where $\bz_{(i)}$ denotes the vector $\bz$ with the $i$th element removed, $j\sim i$ denotes that regions $i$ and $j$ are neighbors, and $m_i$ denotes the number of neighbors for region $i$.

To measure the informativeness of the CAR model framework in~\eqref{eq:bym}, we require an estimate of $\Var\left[\theta_i\given \bbeta, \sigma^2, \tau^2, \btheta_{(i)}\right]$ --- i.e., the variance of $\theta_i$ after integrating out $\bz$ and conditioning on $\btheta_{(i)}$. While $\Var[\theta_i\given \bbeta, \sigma^2, \tau^2, \btheta_{(i)}]$ is a complex function of the full adjacency structure and while~\eqref{eq:bym} and~\eqref{eq:car} can be used directly to obtain a \emph{lower} bound of $\sig^2+\tau^2\slash m_i$ for $\Var\left[\theta_i\given \bbeta, \sigma^2, \tau^2, \btheta_{(i)}\right]$ --- a value that can be obtained as $y_j \to \infty$ --- we follow the guidance of \citet{bym:info} and base our calculations on the \emph{upper} bound of $\Var\left[\theta_i\given \bbeta, \sigma^2, \tau^2, \btheta_{(i)}\right]<\sigma^2+(\sigma^2+\tau^2)/m_i$. Using this bound and the approximation from~\eqref{eq:a} yields a lower bound on the informativeness of the model in~\eqref{eq:bym}, which can be expressed as
\begin{align}
\widehat{a}_i =\frac{1+\exp(\bx_i^T\bbeta)}{\sigma^2+\left(\sigma^2+\tau^2\right)/m_i}-\frac{\exp(\bx_i^T\bbeta)}{1+\exp(\bx_i^T\bbeta)}. \label{eq:info}
\end{align}
Due to the heterogeneity in $m_i$ between counties and a desire to facilitate comparisons of model informativeness between analyses of different datasets, we follow the recommendation of \citet{bym:info} and calculate~\eqref{eq:info} for a county with $m_0=3$ neighbors and use this to quantify the \emph{global} model informativeness, $\widehat{a}_0$. Technical details regarding imposing constraints on $\widehat{a}_0$ within a Markov chain Monte Carlo (MCMC) algorithm are provided in Web Appendix~B.

\section{Simulation Study}\label{sec:sim}
The methodological objective of this paper is to devise an approach for approximating (and/or \emph{controlling}) the informativeness of the CAR model framework of \citet{bym} in the context of binomially distributed count data. In Section~\ref{sec:methods}, we proposed an approximation of the informativeness of logitnormal distribution by first linking it to a beta distribution in the conjugate beta-binomial setting. If this approximation proves to be accurate, the extension from \emph{independent} to \emph{dependent} logitnormal random variables --- e.g., those with the CAR model structure in~\eqref{eq:bym} --- should follow the guidance provided in Section~\ref{sec:bym}, per \citet{bym:info}. Thus, while in Figure~\ref{fig:fig1} we demonstrated that the approximation is quite accurate when there was agreement between the estimates incidence rate from the data and from the prior --- i.e., when $\pi_0 = y_i\slash n_i = a_i\slash \left(a_i+b_i\right)$ --- for selected $\pi_0$ and for a given value of $a_i$,
the goal of this section is to evaluate the performance of the approximation via simulation across a wide spectrum of different scenarios.

$L=100$ sets of data were generated from $y_i \sim \Bin\left(n_i,\pi_i\right)$
with $\pi_i\sim\tbeta\left(a,b\right)$, for $i=1,...,I$, for various numbers of observations, $I$, 
levels of prior informativeness, $a$, 
and prior expected incidence rates, $E\left[\pi_i\right]=\pi_0$. 
For each level of $\pi_0\in \left\{0.01, 0.05, 0.10, 0.20, 0.40\right\}$, sample sizes, $n_i$, were sampled from a uniform distribution such that $E\left[y_i\given n_i,\pi_0\right] \in \left[1,20\right]$.  We then considered levels of $a= \left\{4,8,12,16,20\right\}$ that correspond to a range from
relatively noninformative models to models that have the potential be more informative than all of the observations in the dataset.  To ensure that our results are applicable to a broad range of real-world scenarios, we consider scenarios ranging from having relatively few spatial regions ($I=50$, which is less than the average number of counties per state in the U.S.) to having a large number of spatial regions ($I=200$, which is more than any state in the U.S.\ except Texas) and scenarios with incidence rates ranging from rare ($\pi_0=0.01$) to quite common ($\pi_0 = 0.40$).
For brevity, we restrict our focus \emph{here} to the case where $I=100$ with results for other values of $I$ presented in Web Appendix~C.

To model the data, we used two separate approaches. First, we analyzed the data by effectively using the \emph{true} model --- i.e., $y_i\sim\Bin(n_i,\pi_i)$ with $\pi_i\sim\tbeta(a,b)$ and vague hyperpriors of the form $a\sim\Unif(0,100)$ and $\pi_0=a/(a+b)\sim\Unif(0,1)$. While it should be expected that this model specification would yield \emph{accurate} estimates of $a$ and $\pi_0$, the objective here is to assess the \emph{precision} of these posterior distributions. Next, we analyzed the data using a logitnormal prior specification --- i.e., $y_i\sim\Bin(n_i,\pi_i)$ with $\pi_i\sim\LogitN(\mu,1/\gamma)$ and vague hyperpriors of the form $\mu\sim\Unif(-10,10)$ and $\gamma\sim\Unif(\gamma_L,\gamma_U)$ where $\gamma_L$ and $\gamma_U$ represent bounds based on~\eqref{eq:a} that aim to restrict the model's informativeness to $\widehat{a}\in (0,100)$, mimicking the prior bounds used for $a$ in the beta-binomial model specification. Both models were fit via MCMC sampling using the {\tt rjags} package \citep{rjags} with 20,000 iterations, discarding the first 5,000 iterations as burn-in and thinning samples by a factor of three to reduce autocorrelation, resulting in 5,000 iterations' worth of samples.

\begin{figure}[t]
    \begin{center}
		\subfigure[Model Comparison: $a=12$, $\pi_0=0.1$]{\includegraphics[width=.45\textwidth]{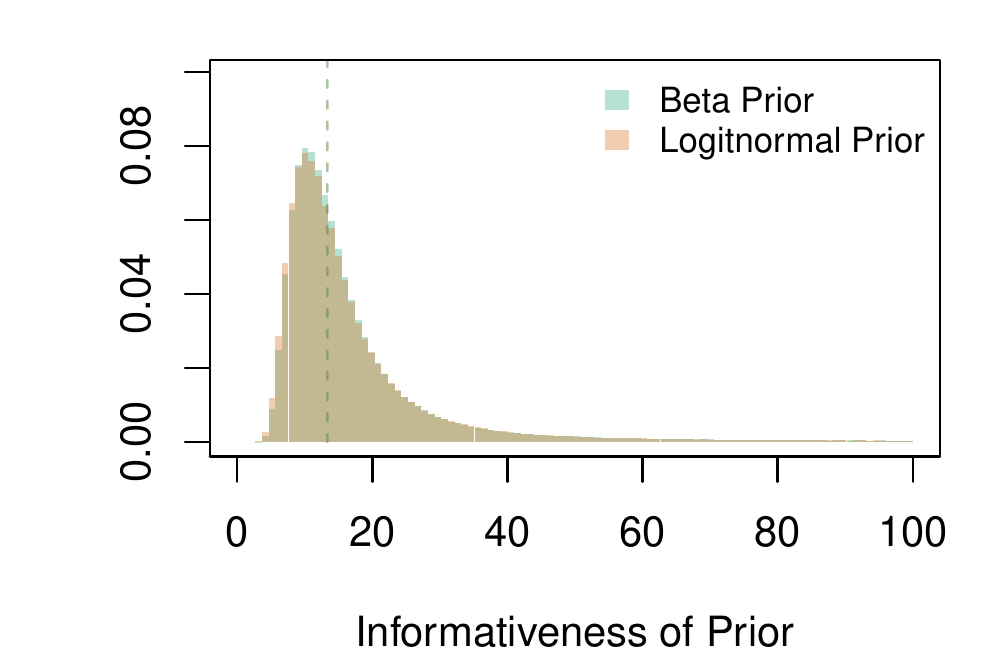}\label{fig:fig21}}
		\subfigure[RMSE: LogitNorm vs.\ Beta]{\includegraphics[width=.4\textwidth]{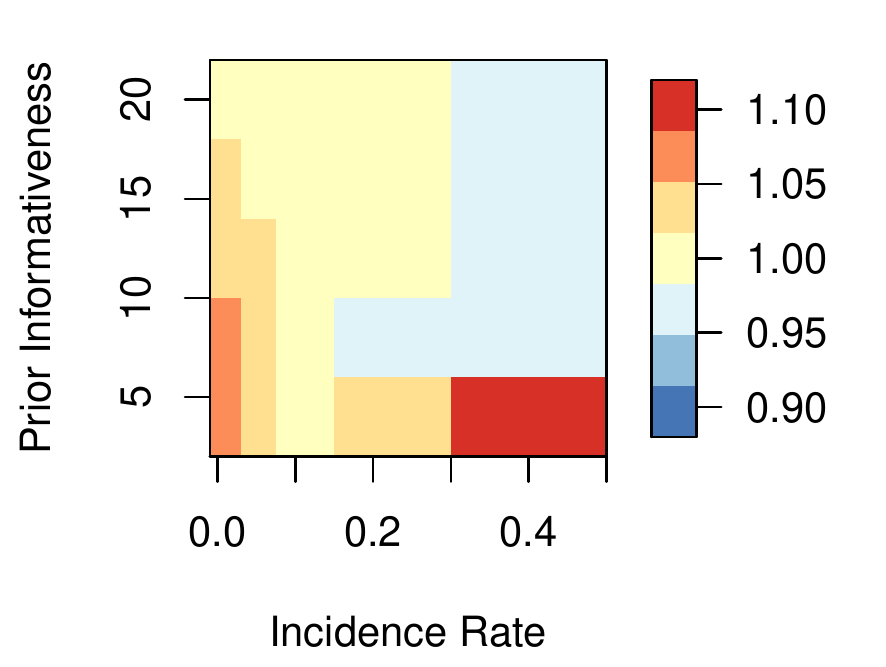}\label{fig:fig22}}
    \end{center}
    \caption{Comparison of the estimated model informativeness from the logitnormal and beta prior specifications.  Panel~(a) compares the posterior distribution for the two models' informativeness parameters under the scenario $a=12$ and $\pi_0=0.1$.  Panel~(b) displays the ratio of the RMSE under the two prior specifications across various levels of $a$ and $\pi_0$, with values greater than 1 indicating better performance under the beta-binomial model.}
	\label{fig:fig2}
\end{figure}

Figure~\ref{fig:fig2} compares the estimated model informativeness under the logitnormal and beta prior specifications. In Figure~\ref{fig:fig21}, we see that when $a=12$ and $\pi_0=0.1$, the posterior distributions for the respective model informativeness parameters --- i.e., $a$ under the beta prior and $\widehat{a}$ from~\eqref{eq:binlogit} under the logitnormal prior --- are practically indistinguishable. Broadly speaking, there is a high degree of agreement between the posterior distributions of the models' informativeness across a range of values of $\pi_0$ and levels of $a$, as demonstrated in {Figures~C.1--C.5 of the Web Appendix}. Evidence of this agreement is summarized in Figure~\ref{fig:fig22}, which displays the ratio of the root mean squared error (RMSE) under the two prior specifications across the various scenarios considered in this study, with values greater than 1 indicating better performance under the beta-binomial model. Here we see that the largest discrepancy between the two approaches occurs when $a$ is small and $\pi_0$ is large --- i.e., low informativeness with high incidence rates. As shown in {Figures~C.1--C.5 of the Web Appendix}, this discrepancy in the RMSEs corresponds to a slight \emph{underestimation} of $a$ under the logitnormal prior (i.e., $E\left[\widehat{a}\given \by\right]=3.44$ on average) compared to a slight \emph{overestimation} under the beta prior ($E\left[a\given \by\right]=4.25$ on average) with nearly identical levels of precision.

\begin{figure}[t]
	\begin{center}
	\subfigure[Model Comparison: $a=12$, $\pi_0=0.1$]{\includegraphics[width=.45\textwidth]{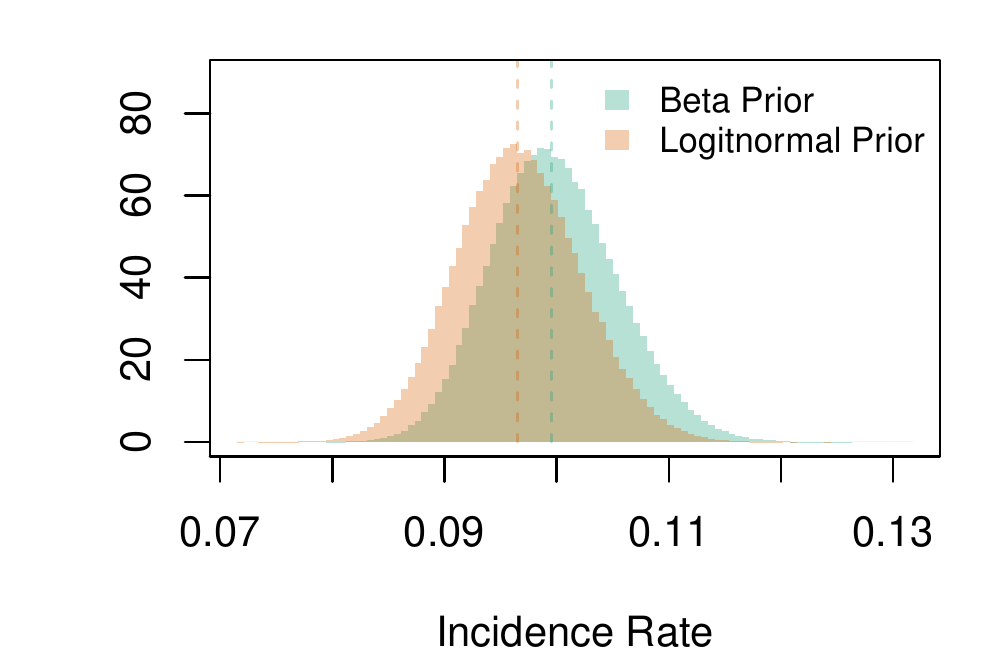}\label{fig:fig31}}
    \subfigure[RMSE: LogitNorm vs.\ Beta]{\includegraphics[width=.4\textwidth]{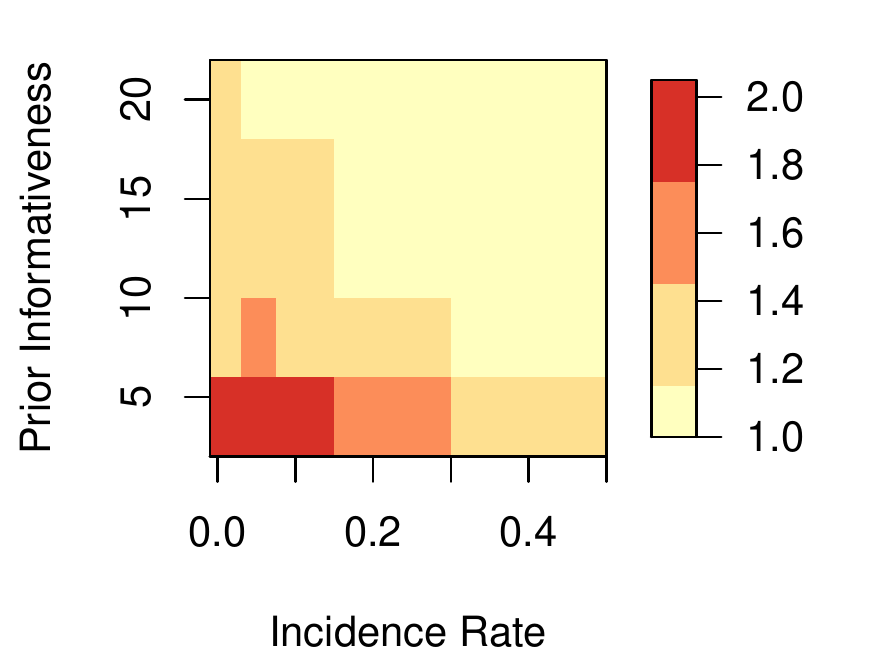}\label{fig:fig32}}
    \end{center}
    \caption{Comparison of the logitnormal and beta prior specification`s estimated incidence rates. Panel (a) compares the posterior distribution of the incidence rate under the scenario where $a=12$ and $\pi_0=0.1$. Panel (b) displays the ratios of the root mean squared error (RMSE) of the two model specifications under the various scenarios considered in this study (values greater than one indicate better performance under the beta prior).}
    \label{fig:fig3}
\end{figure}

While the focus of \emph{this paper} is to quantify the model's informativeness, this work would all be for naught if the logitnormal prior specification produced incorrect inference on the true parameter of interest --- the incidence rate. As illustrated in Figure~\ref{fig:fig31}, the logitnormal prior specification yields a posterior distribution for $\pi_0$ that exhibits a small amount of negative bias when $a=12$ and $\pi_0=0.1$ (i.e., $E\left[\pi_0\given \by\right]=0.097$ on average), while the posterior distribution under the beta prior specification is effectively unbiased.  This is similar to what was observed in Figure~\ref{fig:fig1} and, as shown {Figures~C.1--C.5 of the Web Appendix}, this phenomenon occurs in all of the simulation scenarios studied. Fortunately, as evidenced by the RMSE ratios displayed in Figure~\ref{fig:fig32}, the discrepancies between these two specifications tend to quite small and decrease as $a$ and/or $\pi_0$ increase.

Based on these results and those provided in Web Appendix~C, we conclude that the logitnormal prior specification under~\eqref{eq:res} yields essentially the same inference as the beta prior specification and --- more importantly --- that our measure of the logitnormal prior specification's informativeness is comparable to that of the beta prior specification.

\section{Analysis}\label{sec:analysis}
Having demonstrated the accuracy of the model informativeness estimates derived in Section~\ref{sec:logit}, we now shift our focus to the analysis of the Pennsylvania birth dataset described in Section~\ref{sec:data}.  In particular, we will model the number of low and very low weight births as being binomially distributed out of the total number of births --- stratified by race --- with an aim to estimate geographic and racial disparities in the respective incidence rates in Pennsylvania counties.  Motivated in part by the sparsity of data illustrated in Table~\ref{tab:descriptive}, we will use the CAR model framework of \citet{bym} to model the underlying incidence rates with $\bx_i^T\bbeta=\beta_0$ and vague priors based on the work of \citet{bernardinelli} --- i.e., $p(\beta_0) \propto 1, \sigma^2 \sim \IG(1,1\slash 100)$, and $\tau^2 \sim \IG(1,1\slash 7)$.  After assessing the model's informativeness, $\widehat{a}_0$, per~\eqref{eq:info} for each race compared the amount of data we have, we will reanalyze the data by constraining $\widehat{a}_0$ to be less than 
five low or very low weight births per county and describe how this restriction affects the inference we make.
A detailed analysis of the low birthweight data is provided in Section~\ref{sec:lbw} with additional results and convergence diagnostics in Web Appendix~D,
while a more brief summary of the results for the very low birthweight data is provided in Section~\ref{sec:vlbw} with a more thorough presentation of the results reserved for Web Appendix~E.

\subsection{Low Birth Weight}\label{sec:lbw}
\begin{figure}[t]
    \begin{center}
        \subfigure[Model Informativeness]{\includegraphics[width=.45\textwidth]{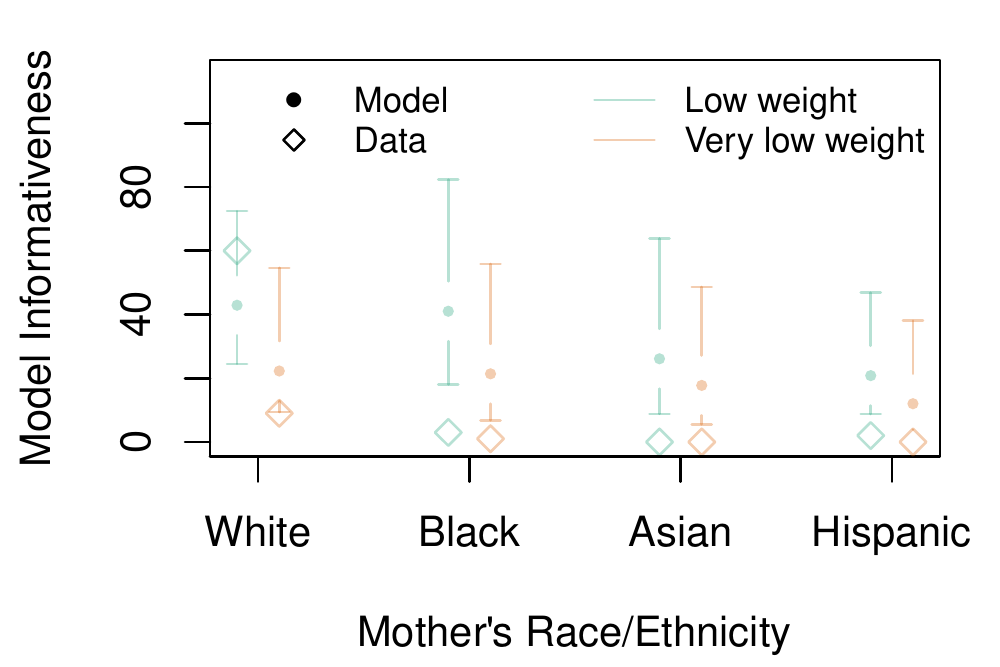}\label{fig:ci}}
        \subfigure[White Mothers in Armstrong County]{\includegraphics[width=.45\textwidth]{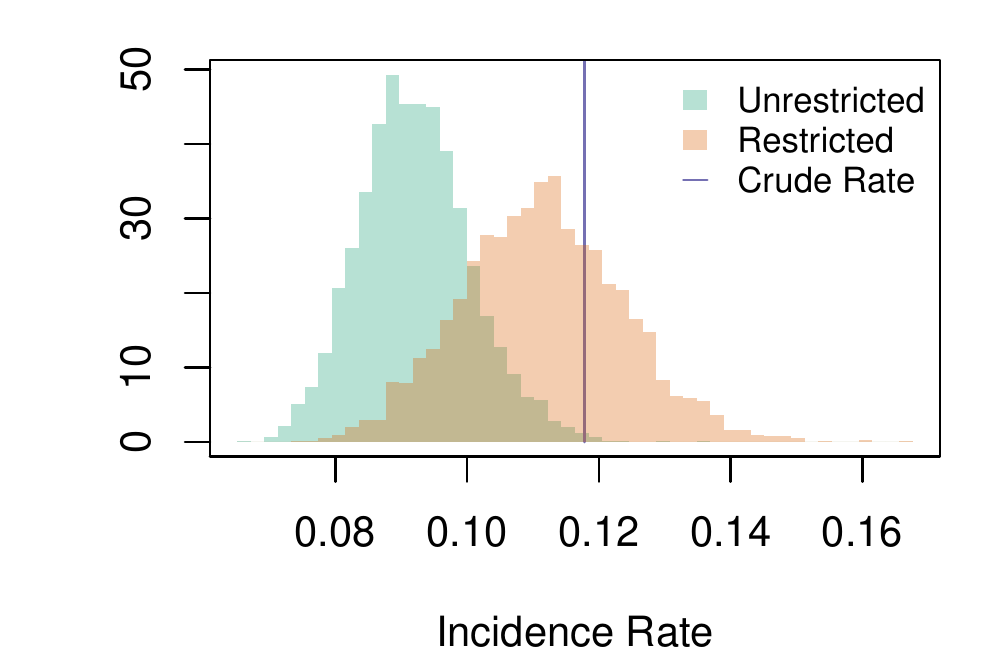}\label{fig:armstrong}}\\
    \end{center}
    \caption{Selected results from the Pennsylvania birth data analyses.  Panel~(a) compares the estimated informativeness of the CAR model specification in~\eqref{eq:bym} by race/ethnicity for the analyses of the low and very low birthweight datasets compared to the median number of low and very low weight births in Pennsylvania counties.  Panel~(b) compares the posterior distribution of the incidence of low birthweight in Armstrong County from the unrestricted and restricted models to the observed crude rate ($70\slash 594= 0.117$).}
	\label{fig:fig6}
\end{figure}

Figure~\ref{fig:ci} compares posterior summaries of the informativeness of the CAR model framework in~\eqref{eq:bym} to the median number of low weight births in Pennsylvania counties by race.  Here, we see clear evidence that the information contributed by the model is overpowering the data for the vast majority of counties, particularly for racial/ethnic minorities.  Specifically, the estimates in Figure~\ref{fig:ci} indicate that the model is contributing the equivalent of over 40 low weight births per county for black mothers and over 20 low weight births for Asian and Hispanic mothers, despite most counties having fewer than 10 low weight births for each of these populations, as shown in Table~\ref{tab:descriptive}.  To see how this affects the estimates we obtain from the model, Figure~\ref{fig:armstrong} shows how the estimated incidence rate for white mothers in Armstrong County was pulled from its observed rate of $70\slash 594= 0.117$ toward the rate of its neighboring counties, 0.071.  Not only does this plot show how the CAR model framework in~\eqref{eq:bym} has the potential to overwhelm and oversmooth estimates, but this is occurring in a region with a relatively large number of low weight births.

\begin{figure}[t]
    \begin{center}
        \subfigure[White Mothers; Unrestricted]{\includegraphics[width=.36\textwidth]{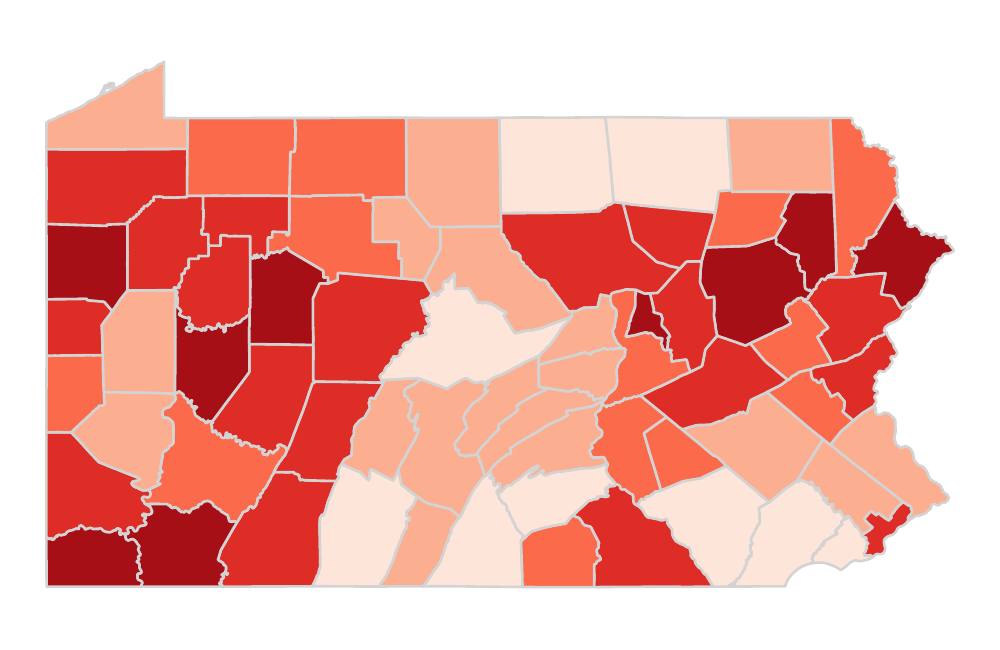}\label{fig:fullwhite}}
        \subfigure[White Mothers; Restricted]{\includegraphics[width=.36\textwidth]{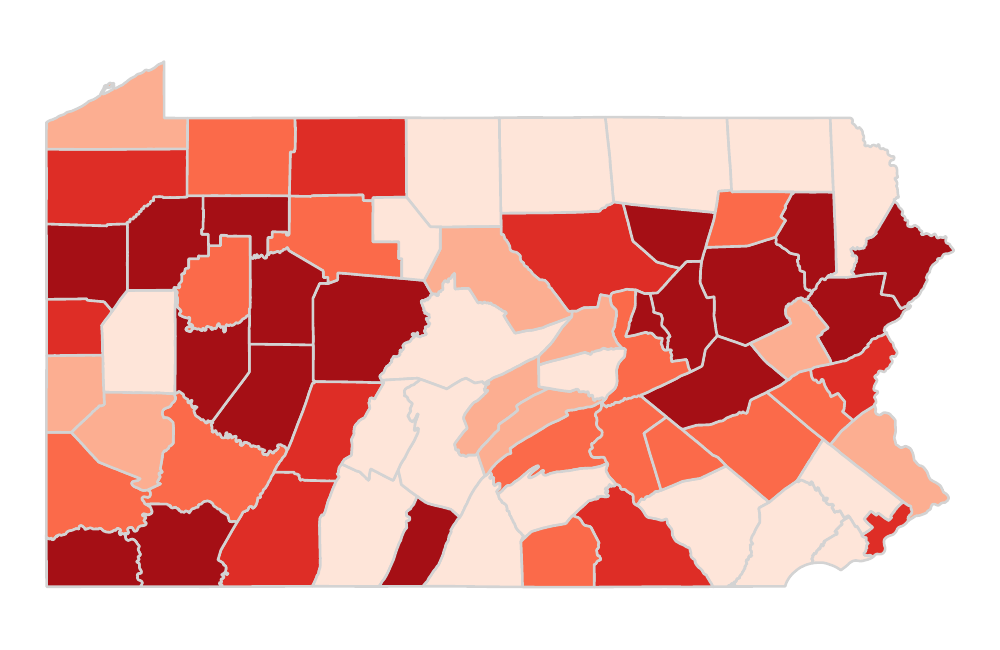}\label{fig:reswhite}}
        \includegraphics[width=.18\textwidth]{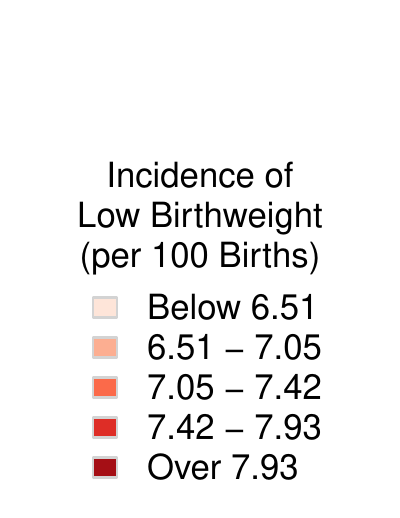}\\
        \subfigure[Black Mothers; Unrestricted]{\includegraphics[width=.36\textwidth]{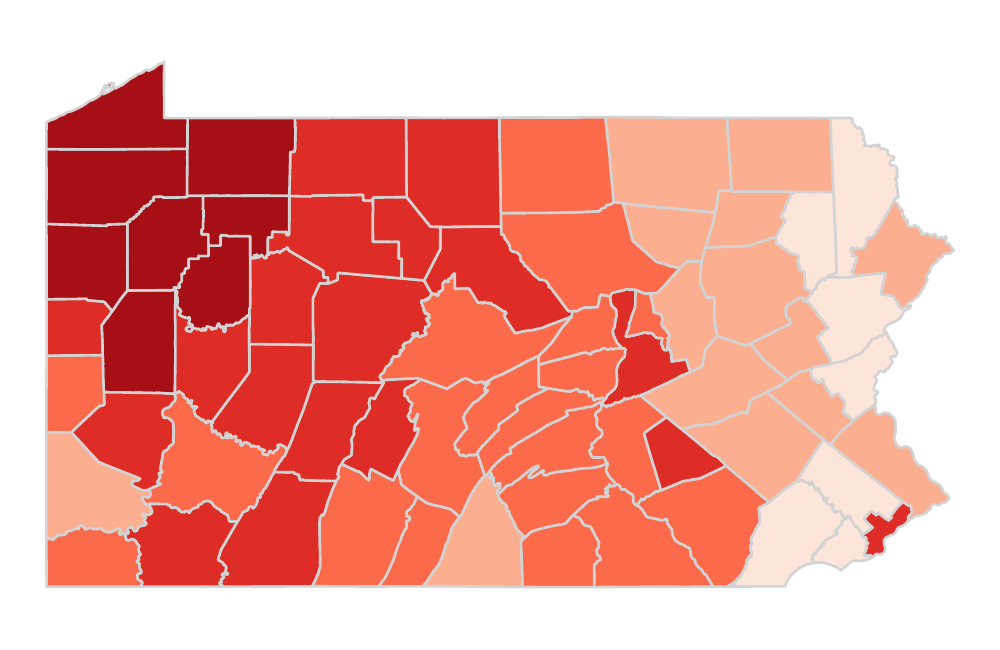}\label{fig:fullblack}}
        \subfigure[Black Mothers; Restricted]{\includegraphics[width=.36\textwidth]{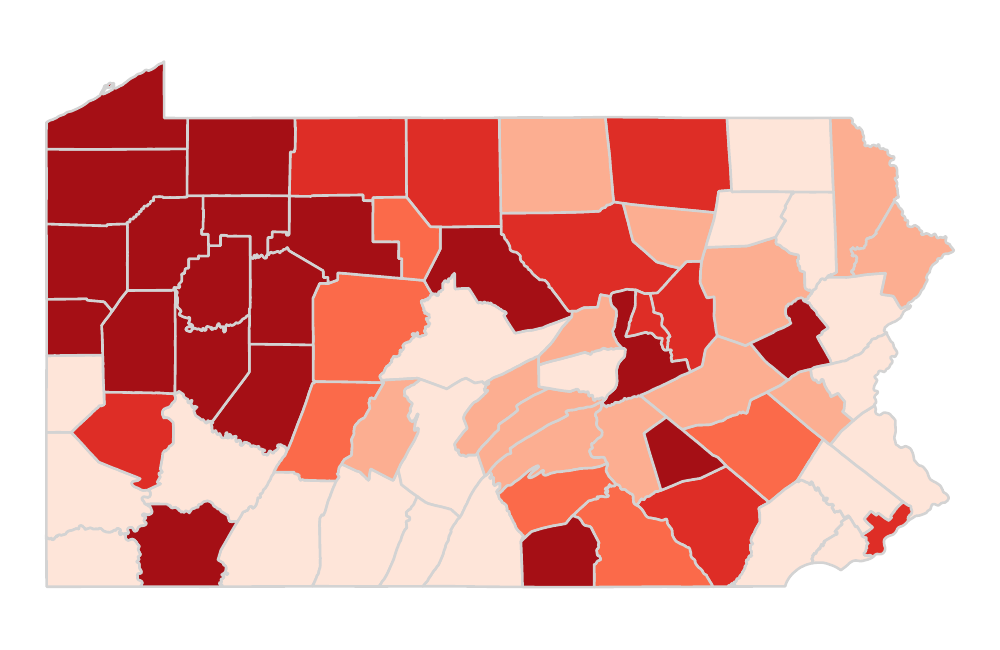}\label{fig:resblack}}
        \includegraphics[width=.18\textwidth]{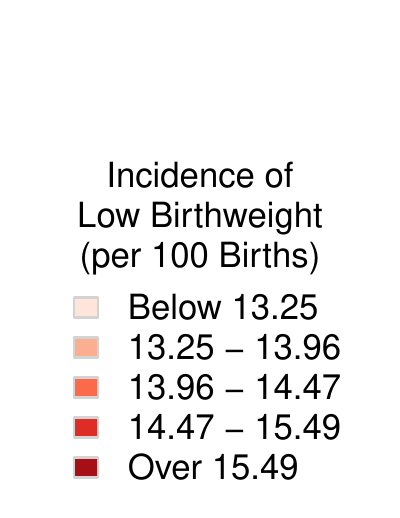}
    \end{center}
    \caption{Maps of incidence of low birthweight for white and black mothers from the unrestricted and restricted models.}
    \label{fig:maps}
\end{figure}

\begin{figure}[t]
    \begin{center}
        \subfigure[B/W Disparities in Select Counties]{\includegraphics[width=.36\textwidth]{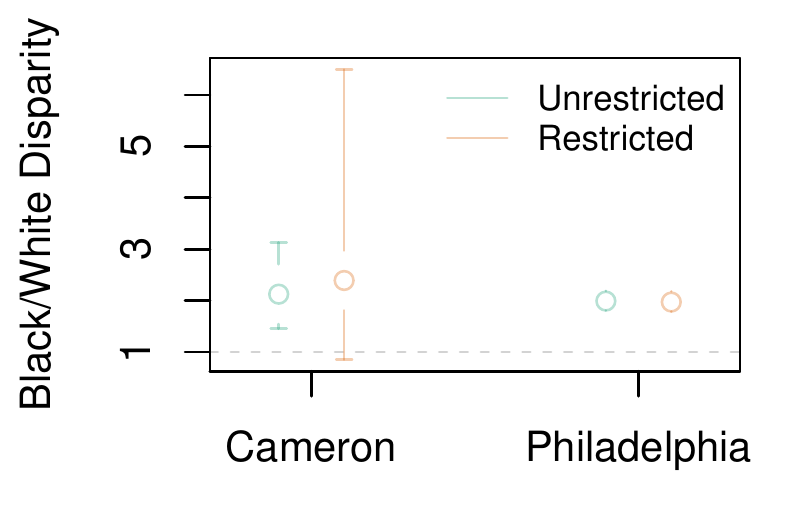}\label{fig:cvp}}
        \subfigure[B/W Disparities; Unrestricted]{\includegraphics[width=.36\textwidth]{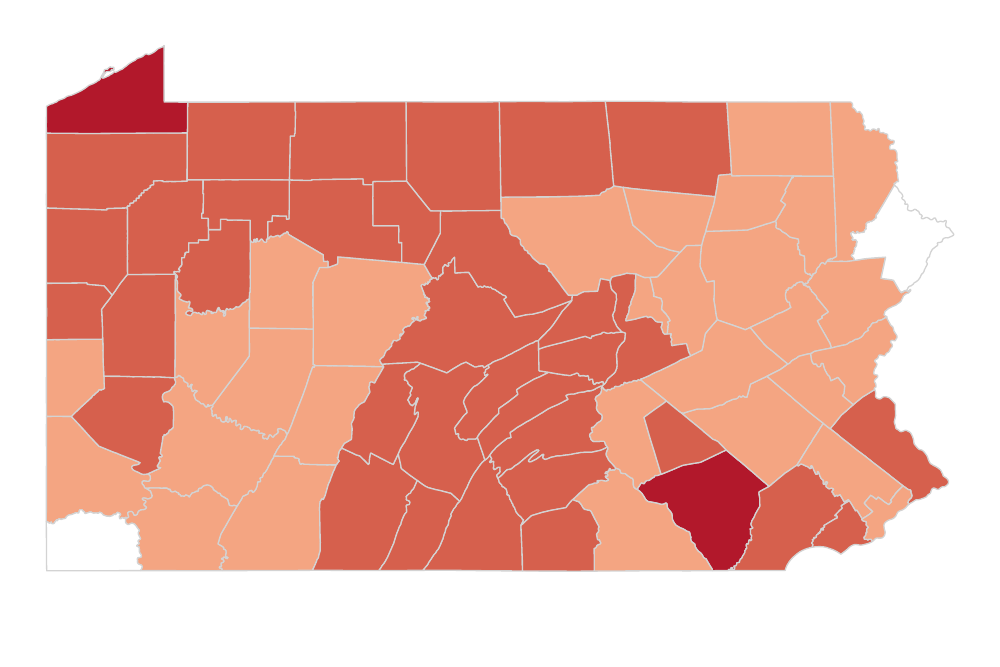}\label{fig:bwmap}}
        \includegraphics[width=.18\textwidth]{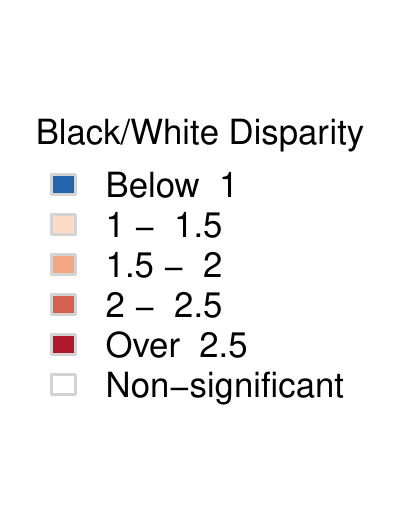}
    \end{center}
    \caption{Investigation of black/white disparities in low birthweight.  Panel~(a) compares the estimated disparities from the two models in one of PA's least populated counties and its most populace county: Cameron and Philadelphia Counties, respectively.  Panel~(b) displays the disparities as estimated by the unrestricted CAR model.}
    \label{fig:bw}
\end{figure}

To illustrate how this oversmoothing affects estimates overall, Figure~\ref{fig:maps} displays maps of the estimated incidence rates for white and black mothers. Here, we see that estimates of the incidence of low birthweight based on the unrestricted CAR model will be largely driven by the CAR model's spatial structure, as observed in the degree of smoothing exhibited in Figures~\ref{fig:fullwhite} and~\ref{fig:fullblack}; similar results are observed for other race/ethnicities in Figures~D.1--D.3 of the Web Appendix.  Furthermore, inference on quantities such as the black/white disparity in incidence rates is also affected, as shown in Figure~\ref{fig:bw}.  Figure~\ref{fig:cvp} highlights that not only are estimates of the black/white disparity ``statistically significant'' in highly populated counties like Philadelphia County, the CAR model in~\eqref{eq:bym} can also produce ``significant'' disparities in sparsely populated rural counties like Cameron County, where just 21 \emph{total} births were observed for white mothers and only \emph{one} birth from a black mother.  All told, the black/white disparity was deemed ``statistically significant'' in all but two counties (Figure~\ref{fig:bwmap}).

In an effort to reduce the amount of spatial smoothing in estimates of the incidence rate and, perhaps more importantly, to discourage pronouncing disparities based on little-to-no data as being ``statistically significant'', we reanalyze the low birthweight data under the constraint that $\widehat{a}_0<5$ for each race/ethnicity.  The effect of this reduction in informativeness is immediately apparent in the estimated incidence of low birthweight for black mothers shown in Figure~\ref{fig:resblack}.  Here, while we still see evidence of spatial smoothing in parts of the state with limited data from black mothers (e.g., the north central region), these rates are allowed to be more consistent with their observed data than be forced to be consistent with the gradient between nearby population centers.  Moreover, while these rates are more stable than the observed rates based solely on the data, their precision remains low enough that the estimated black/white disparity in many of the predominantly white, rural counties in the state are no longer deemed ``statistically significant'', as illustrated in Figure~\ref{fig:cvp}, where the $\widehat{a}_0<5$ constraint causes our estimate of the black/white disparity in Cameron County to have much greater uncertainty.  In contrast, the estimate of the racial disparity in Philadelphia County under the $\widehat{a}_0<5$ constraint is consistent with the estimate from the unrestricted model due to the amount of information contributed by the data.

While not shown here, the degree of spatial smoothing in estimates of the incidence of low birthweight for Asian and Hispanic mothers is similar to those shown for black mothers.  Due to the relative similarity in the overall incidence of low birthweight for white mothers and their Asian and Hispanic counterparts, few counties exhibit disparities that would be deemed ``statistically significant'' under either model.  These results are discussed at greater length in Appendix~D, with maps of the incidence of low birthweight displayed in Figure~D.2 and maps of the racial disparities relative to white mothers shown in Figure~D.4.

\subsection{Very Low Birth Weight}\label{sec:vlbw}
A benefit of analyzing the incidence of both low and very low birthweight is that we can investigate the CAR model's tendency to be overly informative across a broader range of incidence rates.  While race-specific incidence rates for low birthweight ranged from 7\% to over 14.4\% per Table~\ref{tab:descriptive}, the incidence of very low birthweight is less than 3\% for all race/ethnicities studied here.  As a result, investigating the incidence of very low weight births allows us to explore the informativeness of the CAR model in the presence of very limited data.

As shown in Figure~\ref{fig:ci}, the estimated model informativeness in the analysis of the very low birthweight data was less than it was in the analysis of the low birthweight data.  Nevertheless, the CAR model in~\eqref{eq:bym} still contributes the equivalent of between 10--20 very low weight births per county for each of the four race/ethnicities studied, values that exceed the observed counts in all but the largest counties in the state.  As to be expected, this can lead to overly smooth geographic trends in incidence rates, as demonstrated in Figures~E.1--E.3 of the Web Appendix.  It should also be noted that maps of the incidence rates based on the CAR model under the $\widehat{a}_0<5$ restriction also exhibit evidence of potential oversmoothing for Asians and Hispanics.  This, however, should not be too surprising as the vast majority of counties experienced fewer than 5 very low weight births from Asian and Hispanic mothers, and thus even our ``restricted'' model may yet be too informative.  This topic will be more fully discussed shortly.

\section{Discussion}\label{sec:disc}
This paper has extended the framework for measuring and controlling the informativeness of spatial models proposed by \citet{bym:info} to the case of binomially distributed count data.  Unlike when the data are Poisson distributed --- where properties of lognormal random variables could be leveraged to create a model specification comparable to the conjugate gamma prior --- creating a logitnormal model with similar properties to the binomial's conjugate prior required the use of the delta method.  Fortunately, as demonstrated in Section~\ref{sec:sim}, the proposed approximation appears highly accurate for a variety of underlying levels of model informativeness and incidence rates.  This is beneficial because while the Poisson-based framework of \citet{bym:info} was appropriate for rare event data, the binomial framework proposed here is applicable for both rare and common outcomes alike.

While the analysis of binomially distributed count data using spatial models is not itself a novel concept, the primary motivation of this work was to shine light on how informative these sorts of models can be.  This is particularly true when analyzing data stratified by factors such as race/ethnicity, where the sample sizes and the number of events can vary wildly from one region to the next.  In these settings, spatial models without restrictions on their informativeness can easily overwhelm the data from regions with limited data and those which lack racial diversity.  While increasing the precision of our estimates via spatial smoothing is often a motivation for fitting spatial models, models that contribute the equivalent of 40 events per region when the majority of regions have fewer than 10 can produce overconfident and potentially spurious results and should thus be avoided. 

This discussion then leads naturally to the question: \emph{What \emph{is} an acceptable level of model informativeness?}  Not only is this a question for which there is likely no ``standard'' answer, even providing \emph{guidance} is not straightforward.  On one hand, a sensible approach could be to use summary statistics of the observed data to select a threshold for the model's informativeness --- e.g., $\widehat{a}_0$ must not exceed the average number of events per spatial region.  As demonstrated in Section~\ref{sec:analysis}, however, this could lead to \emph{very} different levels of informativeness across factors such as race/ethnicity, a feature that could be less desirable than using a common threshold such as $\widehat{a}_0 < 5$ as a rule-of-thumb.  Moreover, being armed with the capability to set restrictions on the informativeness of the model may in turn provide guidance to researchers on how to \emph{aggregate} their data.  For instance, we may desire to produce estimates of the incidence rate that --- on average --- have a level of precision that could be obtained when $y_i + a_i$
exceeds a certain threshold.  From this, users could then determine the number of years of data that would be required to be combined in order for that threshold to be exceeded in a majority of counties for a given level of $a_i$.

Finally, we'd be remiss to not acknowledge that alternatives to the CAR model framework of \citet{bym} exist and have gained popularity, including the Leroux CAR model of \citet{leroux:car} and the directed acyclic graph autoregressive model of \citet{datta:dagar}.  To extend this approach to these alternative methods, bounds on their analogous $\Var\left[\theta_i\given \bbeta, \sigma^2, \tau^2, \btheta_{(i)}\right]$ expressions would be required but could then combined with the approximation from~\eqref{eq:a} to establish the appropriate measures of model informativeness.  That said, we do not expect that these alternative approaches would be \emph{immune} from the issue of oversmoothing, thus our future work includes exploring extensions of this nature.

\bibliographystyle{jasa}
\bibliography{bin_arxiv}

\begin{thebibliography}{19}
\newcommand{\enquote}[1]{``#1''}
\expandafter\ifx\csname natexlab\endcsname\relax\def\natexlab#1{#1}\fi

\bibitem[\protect\citename{Barker et~al., }1995]{Barker}
Barker, D., Osmond, C., and Simmonds, S. (1995).
\newblock \enquote{Fetal and placenta size and risk of hypertension in adult
  life.}
\newblock {\em BMJ\/}, 301, 259--62.

\bibitem[\protect\citename{Bernardinelli et~al., }1995]{bernardinelli}
Bernardinelli, L., Clayton, D., and Montomoli, C. (1995).
\newblock \enquote{Bayesian estimates of disease maps: How important are
  priors?}
\newblock {\em Statistics in Medicine\/}, 14, 2411--2431.

\bibitem[\protect\citename{Besag et~al., }1991]{bym}
Besag, J., York, J., and Molli\'e, A. (1991).
\newblock \enquote{Bayesian image restoration, with two applications in spatial
  statistics.}
\newblock {\em Annals of the Institute of Statistical Mathematics\/}, 43,
  1--59.

\bibitem[\protect\citename{Botella-Rocamora et~al., }2015]{b-r:2015}
Botella-Rocamora, P., Martinez-Beneito, M.~A., and Banerjee, S. (2015).
\newblock \enquote{A unifying modeling framework for highly multivariate
  disease mapping.}
\newblock {\em Statistics in Medicine\/}, 34, 1548--1559.

\bibitem[\protect\citename{Brillinger, }1986]{brillinger}
Brillinger, D.~R. (1986).
\newblock \enquote{The natural variability of vital rates and associated
  statistics.}
\newblock {\em Biometrics\/}, 42, 693--734.

\bibitem[\protect\citename{Collins and David, }2009]{collins}
Collins, J. and David, R. (2009).
\newblock \enquote{Racial disparity in low birth weight and infant mortality.}
\newblock {\em Clinics in Perinatology\/}, 36, 63--73.

\bibitem[\protect\citename{Datta et~al., }2019]{datta:dagar}
Datta, A., Banerjee, S., Hodges, J.~S., and Gao, L. (2019).
\newblock \enquote{Spatial disease mapping using directed acyclic graph
  auto-regressive {(DAGAR)} models.}
\newblock {\em Bayesian Analysis\/}, 14, 1221--1244.

\bibitem[\protect\citename{Gelfand and Vounatsou, }2003]{gelfand:mcar}
Gelfand, A.~E. and Vounatsou, P. (2003).
\newblock \enquote{Proper multivariate conditional autoregressive models for
  spatial data analysis.}
\newblock {\em Biostatistics\/}, 4, 11--25.

\bibitem[\protect\citename{Goldfarb et~al., }2018]{goldfarb}
Goldfarb, S.~S., Houser, K., Wells, B.~A., Brown~Speights, J.~S., Beitsch, L.,
  and Rust, G. (2018).
\newblock \enquote{Pockets of progress amidst persistent racial disparities in
  low birthweight rates.}
\newblock {\em PLoS ONE\/}, 13, e0201658.

\bibitem[\protect\citename{Hack et~al., }1995]{Hack}
Hack, M., Klein, N.~K., and Taylor, H.~G. (1995).
\newblock \enquote{Long-term developmental outcomes of low birth weight
  infants.}
\newblock {\em Future Child\/}, 5(1), 176--196.

\bibitem[\protect\citename{Leroux et~al., }2000]{leroux:car}
Leroux, B.~G., Lei, X., and Breslow, N. (2000).
\newblock \enquote{Estimation of disease rates in small areas: {A} new mixed
  model for spatial dependence.}
\newblock In {\em Statistical Models in Epidemiology, the Environment, and
  Clinical Trials\/}, eds. M.~E. Halloran and D.~Berry,  179--191. New York,
  NY: Springer New York.

\bibitem[\protect\citename{Martin et~al., }2021]{births:2019}
Martin, J.~A., Hamilton, B.~E., Osterman, M. J.~K., and Driscoll, A.~K. (2021).
\newblock \enquote{Births: {F}inal data for 2019.}
\newblock {\em National Vital Statistics Reports\/}, 70, 1--51.

\bibitem[\protect\citename{{Pennsylvania Department of Health}, }2020]{eddie}
{Pennsylvania Department of Health} (2020).
\newblock
  \url{https://www.health.pa.gov/topics/HealthStatistics/EDDIE/Pages/EDDIE.aspx#}.

\bibitem[\protect\citename{Plummer, }2016]{rjags}
Plummer, M. (2016).
\newblock {\em rjags: {B}ayesian Graphical Models using {MCMC}\/}.
\newblock R package version 4-6.

\bibitem[\protect\citename{Quick et~al., }2015]{hcar}
Quick, H., Carlin, B.~P., and Banerjee, S. (2015).
\newblock \enquote{Heteroscedastic conditional auto-regression models for
  areally referenced temporal processes for analysing {C}alifornia asthma
  hospitalization data.}
\newblock {\em Journal of the Royal Statistical Society, Series C\/}, 64,
  799--813.

\bibitem[\protect\citename{Quick et~al., }2021]{bym:info}
Quick, H., Song, G., and Tabb, L. (2021).
\newblock \enquote{Evaluating the informativeness of the
  {B}esag-{Y}ork-{M}olli\'e {CAR} model.}
\newblock {\em Spatial and Spatio-temporal Epidemiology\/}, 37, 100420.

\bibitem[\protect\citename{Rich-Edwards et~al., }1999]{Rich}
Rich-Edwards, J., Colditz, G., Stampfer, M., Willett, W.~C., Gillman, M.~W.,
  Hennekens, C.~H., Speizer, F.~E., and Manson, J.~E. (1999).
\newblock \enquote{Birthweight and the risk for type 2 diabetes in adult
  women.}
\newblock {\em Annals of Internal Medicine\/}, 130, 278--84.

\bibitem[\protect\citename{Waller et~al., }1997]{waller:carlin}
Waller, L.~A., Carlin, B.~P., Xia, H., and Gelfand, A.~E. (1997).
\newblock \enquote{Hierarchical spatio-temporal mapping of disease rates.}
\newblock {\em Journal of the American Statistical Association\/}, 92,
  607--617.

\bibitem[\protect\citename{{World Health Organization}, }2014]{who}
{World Health Organization} (2014).
\newblock \enquote{{Global Nutrition Targets 2025: Low birth weight policy
  brief}.}
\newblock
  \url{https://apps.who.int/iris/bitstream/handle/10665/149020/WHO_NMH_NHD_14.5_eng.pdf?ua=1}.

\end{thebibliography}

\end{document}